\documentstyle[prl,aps,epsf]{revtex}  

\newcommand{\bra}[1]{\langle{#1}|}
\newcommand{\ket}[1]{|{#1}\rangle}
\newcommand{\braket}[2]{\langle{#1}|{#2}\rangle}

\newcommand{\op}[1]{\hat{#1}}
\newcommand{\adjop}[1]{\op{#1}^\dagger}

\newcommand{\cep}[1]{e^{i{#1}}}
\newcommand{\cem}[1]{e^{-i{#1}}}

\newcommand{\set}[1]{\left\{{#1}\right\}}

\makeatletter
\def\tr{\mathop{\operator@font tr}\nolimits}
\makeatother
\makeatletter
\def\mod{\mathop{\operator@font mod}\nolimits}
\makeatother

\newcommand{\twopiN}{\frac{2\pi}{N}}
\newcommand{\floq}{\chi_q}
\newcommand{\flop}{\chi_p}
\newcommand{\diff}[1]{\,d{#1}}

\newlength{\figheight}
\newlength{\figwidth}
\begin{document}

\title{Decoherence for classically chaotic quantum maps}

\author{Pablo Bianucci$^1$, Juan Pablo Paz$^1$ and Marcos Saraceno$^2$} 
  
\address{$^1$Departamento de F\'{\i}sica ``J.J. Giambiagi'',  
FCEN, UBA, Pabell\'on 1, Ciudad Universitaria, 1428 Buenos Aires, Argentina}  
  
\address{$^2$Unidad de Actividad F\'{\i}sica, Tandar, CNEA  
Buenos Aires, Argentina}  
  
\maketitle  
  
\begin{abstract}  
{We study the behavior of an open quantum system, with an $N$--dimensional
space of states, whose density matrix evolves
according to a non--unitary map defined in two steps: A unitary step, 
where the system evolves with an evolution operator obtained by quantizing 
a classically chaotic map (baker's and Harper's map are the two examples we 
consider). A non--unitary step where the evolution operator for the density
matrix mimics the effect of diffusion in the semiclassical (large $N$) limit. 
The process of decoherence and the transition from quantum to classical 
behavior are analyzed in detail by means of numerical and analitic tools. 
The existence of a regime where the entropy grows with a rate which is
independent of the strength of the diffusion coefficient 
is demonstrated. The nature of the processes that determine the 
production of entropy is analyzed. }
\end{abstract}

\date{\today}  
  
%
%


\draft

\section{Introduction}

Decoherence has been recognized in recent years 
as one of the main ingredients needed to understand the origin of 
the classical world from the fundamental quantum laws 
\cite{ZurekPT,PazZurek00}. Decoherence is a process whose origin is 
conceptually simple: It is due to the
entanglement between the system and its environment 
that is created in the course of their interaction. As a consequence, the 
environment keeps a record of the state of the system, that loses its
purity. Only a small subset of all possible states of the system 
(the so--called pointer states) are relatively stable against 
the interaction. They are the ones which are less likely to become 
entangled with the environment. In the vast majority of cases, 
when the state is a superposition of pointer states, 
the information initially stored in 
the state of the system can never be restored since it irreversibly 
flows into the correlations with the environment. A basic question one
should ask in this context is how fast does the information flow away from
the system. This can be studied, for example, by analyzing the 
evolution of the entropy obtained from the reduced density matrix 
of the system. This has been done for a variety of cases 
(see \cite{PazZurek00}
for a review) and it has been recognized that this process has 
unique features if the system has a classically chaotic counterpart. 
In fact, as conjectured in 
\cite{ZurekPaz94,ZurekPaz95}, the rate of entropy production in 
such cases has a regime that is independent of the strength of the 
coupling between the system and the environment and is entirely 
determined by the dynamical parameters characterizing the chaotic 
evolution. This conjecture was analyzed in the literature mostly 
using numerical tools \cite{MP00,MP01,Zurek00,HuSh,Sarkar,Pattan,Nag}.
In this paper we will present a study of this problem for
some systems that are simple enough to enable both a rigorous numerical
treatment and some analytic estimates. We consider here a quantum 
system with a finite dimensional Hilbert space ($N$ is the number of
dimensions, and we are interested in learning about the behavior 
of the system in the large $N$ limit). For such system we define 
an evolution operator for the density matrix in two steps as follows: first we 
consider a purely unitary evolution defining an operator $U$ which
is such that it corresponds (in the large $N$ limit) to a classically 
chaotic map. 
Then, we define a non--unitary map for the density matrix in such 
a way that (again, in the large $N$ limit) it mimics the effect of
the interaction with an environment producing the same effects one
observes for a Brownian particle (namely, diffusion). For such system, 
we developed a numerical code enabling us to efficiently 
evolve the density matrix  
and study in particular the evolution of the entropy. Our aim 
is to present solid numerical evidence supporting the 
conjecture presented in \cite{ZurekPaz94,ZurekPaz95} and, by combining
the numerical work with analytic calculations, develop new 
intuition on the main processes contributing to entropy production 
for this kind of systems. 

The paper is organized as follows: In Section II we review some basic elements
of the theory of quantum maps. We focus, as in the rest of the paper, on two
specific examples: baker's map (the paradigmatic example of a fully chaotic 
system) and Harper's map (an example of the wide class of kicked maps with 
mixed phase space). There 
are no new results presented in this section and the reader with experience
in the theory of quantum maps can easily skip it. In Section III we describe 
in detail the model for decoherence that we study in this paper. In Section
IV we present the main results concerning the behavior of the entropy as a 
function of time and the evolution of the Wigner function. Finally, in Section 
IV we present our conclusions. An Appendix contains  technical details about
the phase space representation we use in this paper (the discrete Wigner 
function).

\section{Quantum Maps}

The construction of the quantum analogue of a classical map follows two well 
defined steps : a kinematical
one, where the nature of the Hilbert space is defined in relationship with
a specific phase space structure, and a dynamical one, where a unitary 
operator defines the evolution for a finite time step. For the study of chaotic
behaviour, a finite phase space is required and therefore the question 
of boundary conditions arise. The simplest case is the torus, where 
periodic boundary conditions are assumed for both the coordinate and 
momentum representations.  The most general quasi-periodic boundary 
conditions are 
\begin{eqnarray}
\braket{q+1}{\psi} & = & e^{i2\pi \floq}\braket{q}{\psi}  \\
\braket{p+1}{\psi} & = & e^{-i2\pi \flop}\braket{p}{\psi}
\end{eqnarray}
where $\floq$ and $\flop$ are fixed, arbitrary real numbers between $0$ and $1$
($2\pi\floq$ and $2\pi\flop$ are called Floquet angles). These
conditions result in a finite dimensional Hilbert space. This space's
dimension, $N$, is related to $\hbar$ by the relation
\begin{equation}
2\pi\hbar N = 1
\end{equation}
which signifies that phase space (of area equal to unity ) is spanned by $N$
states (of area $2\pi\hbar$).
The position and momemntum eigenvalues in this finite dimensional space are
\begin{eqnarray}
\label{eq:q_discr_eigen}
\ket{q_n} & = & \ket{\frac{n+\flop}{N}}, \quad n=0,1,\dots,N-1, \\
\ket{p_m} & = & \ket{\frac{m+\floq}{N}}, \quad m=0,1,\dots,N-1.
\end{eqnarray}
These position and momentum eigenstates are related by a discrete Fourier
transform:
\begin{equation}
\braket{p_m}{q_n} = \frac{1}{\sqrt{N}}e^{-i\twopiN (m+\floq)(n+\flop)}
               \equiv \left( G_N^{\floq,\flop}\right)_{mn}
\label{fourier}	       
\end{equation}

The values of $\floq$ and $\flop$ specify different Hilbert spaces. In the
present paper we will use $\floq = \flop = \frac{1}{2}$, corresponding to
antiperiodic boundary conditions. Cyclic shifts on these two bases are 
implemented by unitary operators \cite{Schwinger}
\begin{eqnarray}
\label{eq:schwinger_ops}
\op{\cal U}\ket{q_n} & = & \ket{q_{n+1}} \\
\op{\cal V}\ket{p_m} & = & \ket{p_{m+1}}.
\end{eqnarray}
These operators and their powers can be combined to produce unitary 
displacement operators in phase space 
\begin{equation}
\op{D}(\Delta p,\Delta q)=
 \op{\cal U}^{\Delta q}\op{\cal V}^{\Delta p}\cep{\frac{\pi}{N}
\Delta p\Delta q}
 \label{eq:D_operators}
\end{equation} 
for integer values of $\Delta p, \Delta p$.
These are the analogues of  the Weyl displacement operators in the continuous
case. Phase space in this context is then assimilated to a discrete 
$N\times N$ grid which will be useful for the representation of quantum 
effects in phase space (see Appendix).

\subsection{The baker's map}
The baker's map is one of the simplest systems displaying strongly chaotic 
behaviour. In spite of its 
simplicity, it has a very rich dynamical behaviour, both in its classical and
its quantum versions. The map is an area preserving
transformation defined in the $[0,1] \times [0,1]$ phase-space square as:
\begin{eqnarray}
 \label{eq:bq}
 q' & = & 2q-[2q]                   \\
 p' & = & \frac{1}{2}(p+[2q])
 \label{eq:bp}
\end{eqnarray}
where the square brackets symbolize the integer part of a real number.
This evolution has a very simple geometrical interpretation, as a "stretching"
step followed by ``cutting'' step, as a baker rolling dough. The map is 
uniformly hyperbolic, with a single Lyapunov exponent, with a value of $\ln 2$.
Moreover, at every point the stable and unstable manifolds are parallel to the
coordinate axes.
The baker's map has a remarkably simple symbolic dynamics, and can be mapped
into an unrestricted Bernoulli shift on two symbols \cite{Arnold-Avez}. If 
the phase space coordinates $p$ and $q$ are written in binary notation
\begin{eqnarray}
q = 0.\epsilon_0\epsilon_1\epsilon_2\epsilon_3\dots   \quad\quad
p = 0.\epsilon_{-1}\epsilon_{-2}\epsilon_{-3}\epsilon_{-4}\dots  \nonumber \\
(p,q) = \dots\epsilon_{-4}\epsilon_{-3}\epsilon_{-2}\epsilon_{-1}\bullet
        \epsilon_0\epsilon_1\epsilon_2\epsilon_3\dots
\end{eqnarray}
The map's action on these symbols is to move the most significant bit of $q$
to $p$, shifting to the right the decimal point in $q$:

\begin{eqnarray}
q' = 0.\epsilon_1\epsilon_2\epsilon_3\dots            \quad\quad
p' = 0.\epsilon_0\epsilon_{-1}\epsilon_{-2}\epsilon_{-3}\dots    \nonumber \\
(p',q') = \dots\epsilon_{-4}\epsilon_{-3}\epsilon_{-2}\epsilon_{-1}
          \epsilon_0\bullet\epsilon_1\epsilon_2\epsilon_3\dots
\end{eqnarray}
Thus each doubly infinite sequence of binary digits represents a unique 
trajectory. The phase space points on this trajectory are obtained by
placing the dot somewhere (the present) and reading off the coordinate 
and momentum to the right and left of it.

\begin{figure}
\center
\epsfxsize=\figwidth
\epsfbox{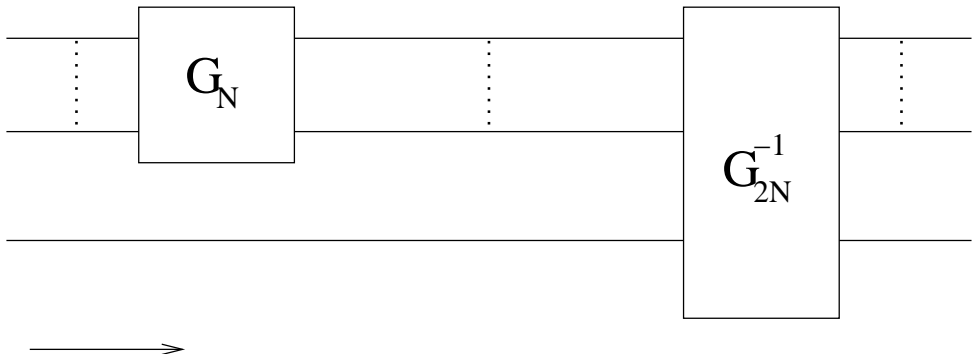}
\vspace {0.25cm}  
\label{fig:bak_circ}
\caption{Quantum circuit which perform a Bernoulli shift, i.e. an iteration of
the quantum baker's map}
\end{figure}

The quantization procedure for the map is not unique but follows closely 
semiclassical
prescriptions. As originally formulated \cite{Balasz-Voros} it used periodic
boundary conditions ($\floq = \flop = 0 $) but was later modified to 
antiperiodic conditions ($\floq = \flop = \frac{1}{2}$)
\cite{Saraceno:Ann.Phys}.
The full semiclassical theory has been developed in \cite{Saraceno:PhysD}. 
The resulting unitary matrix for one step of the map is
\begin{equation}
 \op{B}_{baker} = \op{G}_{2N}^{-1}  \left[ \begin{array}{cc}
                              \op{G}_{N} &      0     \\
                                   0     & \op{G}_{N} \\
                          \end{array} \right]
\end{equation}
A different approach, leading to the same result, was proposed 
in \cite{Schack} (see also \cite{Caves-Schack}) 
interpreting the map as the quantization of a
Bernouilli shift showing that it could be implemented by
elementary gates and thus would be an interesting candidate to be 
run as an algorithm in a quantum
computer. This approach  is strongly based on the symbolic dynamics of the
map and runs as follows: as the position and momentum basis are related 
by a Fourier transform, we can implement the Bernoulli shift - which 
shifts the most significant bit of the coordinate to the most significant 
bit of momentum - using the quantum circuit shown in figure 
\ref{fig:bak_circ}.

The action of this circuit can be understood as follows: the qubits 
on the input
codify in the usual way the eigenvalues of the position operator with the most
significant bit at the bottom; after applying the split
Fourier transform, the most significant qubit now represents the most 
significant
bit of the eigenvalue of the momentum operator. As a final step, an inverse
Fourier transform allows us to look at the final state in the position basis.
This is exactly the circuit representation of the matrix  $\op{B}_{baker}$ .
As is well known \cite{Chuang-Nielsen} the Fourier matrices can be 
further decomposed into elementary gates leading to a circuit representation 
in terms of operations on qubits.

\subsection{The Harper's map}

The baker's map does not capture the full complexity of chaotic motion in 
Hamiltonian systems. The fact that it has uniform hyperbolicity and very
simple manifolds are not very generic properties as the most common situation
is that of a complex mixture of elliptic islands interspersed by chaotic 
regions
with locally defined Lyapounov exponents. To address this more general
situation a different family of maps can be devised whose characteristic is
to alternately produce ``kicks'' of potential or kinetic energy. The combined
action of these kicks is equivalent to that of a periodic time dependent 
Hamiltonian and the resulting motion is area preserving and can mimic the full
complexity of a generic system. A wide variety of maps on the sphere (kicked
tops), on the cylinder (kicked rotors), or on the torus (Harper's map) 
have been extensively studied in the quantum chaos literature 
\cite{Haake,Harper}. 

Here we have studied a specific kicked map, known as the Harper's map 
because of its relationship to the Harper's hamiltonian in solid state 
physics. This
map acts on the unit square with periodic boundary conditions in phase-space
and is defined by the transformation \cite{Harper}
\begin{eqnarray}
q' & = & q - \gamma \sin(2\pi p)    \quad (\mod 1)       \\
p' & = & p + \gamma \sin(2\pi q')   \quad (\mod 1).
\end{eqnarray}
The behaviour of the map is determined by the real parameter $\gamma$. If 
$\gamma\ll 1$ the map approaches an infinitesimal
transformation with approximately conserved energy and its evolution 
is regular. If, on the other hand, $\gamma\approx 1$ the map becomes 
fully chaotic.  In between, the motion presents
the complex mixture of regular and chaotic motion characteristic of 
most realistic systems.

The reason that kicked maps are so popular is that they are very easy to
quantize. In fact the potential and the kinetic kicks are respectively 
diagonal in the coordinate and momentum basis and therefore the full 
evolution consists of these diagonal kicks interspersed by the Fourier 
transformation between these two bases. One full step of the map is then 
the evolution operator $\op{U}$,
\begin{equation}
\op{U} = \op{U}_q\adjop{G}_N  \op{U}_p\op{G}_N
\end{equation}

 For the Harper's map the two diagonal operators $\op{U_q}$ and $\op{U}_p$ 
are

\begin{eqnarray}
\op{U}_q \ket{q_n} & = &
                 \cem{\gamma N \cos (\twopiN ( n + \flop))} \ket{q_n} \\
\op{U}_p \ket{p_k} & = &
                 \cem{\gamma N \cos (\twopiN ( k + \floq))} \ket{p_k}  
\end{eqnarray}

and $\op{G_N}$ is given by (\ref{fourier}).

\begin{figure}
\center
\epsfxsize=\figwidth
\epsfbox{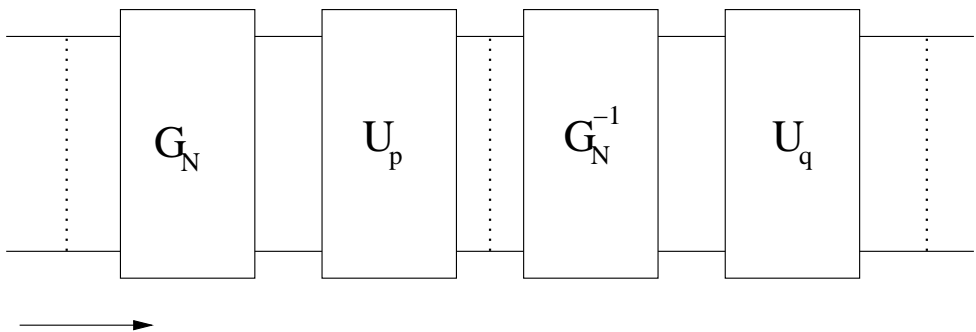}
\vspace {0.25cm}
\label{fig:kick_circ}
\caption{Quantum circuit which implements a quantum kicked map.}
\end{figure}

Kicked maps also lead very naturally to a circuit interpretation, 
like in the case
of the baker's map. The diagonal interactions are controlled phases that 
act among all the qubits while the Fourier transformations can again 
be decomposed into elementary gates. Some of these maps 
(the cat maps, for example) can be efficiently decomposed in terms 
of elementary operatios and have recently been studied as possible 
candidates to be simulated in a quantum computer \cite{Shepelyansky}. 
Figure \ref{fig:kick_circ} shows the structure of a quantum circuit 
which implements a generic kicked map.

\section{Dissipative maps}

Our aim is to study the impact of the process of  
decoherence induced by the interaction between our system and an 
external environment. The system would 
otherwise evolve according to one of the unitary operators described 
in the previous section. In general, modeling the coupling to 
the outside world may be complicated. Here, we will not use a  
microscopic model of this interaction but will describe the effect
of the environment in a phenomenological way by defining a dissipative map
for the evolution of the density matrix of the system. 
As decoherence generally induces a loss of purity, the first 
important point to notice is that the state of the system should be 
defined in terms of a density matrix $\rho$.
In the absence of any coupling to the environment, the density matrix
$\rho$ evolves unitarily according to the map 
\begin{equation}
\op{\rho}'=\op{U}\op{\rho}\adjop{U}.\label{unitarymap}
\end{equation}
The coupling to the environment would induce nonunitary evolution. 
To correspond to an allowed temporal evolution (that should 
come from a unitary map for the whole Universe), the nonunitary map 
for the density matrix has to satisfy some constraints. 
Assuming that: i) it is linear and preserves hermiticity, ii) it 
is trace preserving and
iii) it preserves the complete positivity of the density matrix
\cite{Chuang-Nielsen}; the map is strongly constrained. Moreover, 
if one imposes also Markovian behavior (neglecting all memory effects) one
can show that the most general map for the density matrix of the system should be of the form:
\begin{equation}
\op{\rho}' = \sum_k \op{E}_k \op{\rho} \adjop{E}_k
\label{eq:opsum}
\end{equation}
Equation (\ref{eq:opsum}) is known as the operator sum representation (or Kraus
representation) of the superoperator which maps $\op{\rho}$ onto $\op{\rho}' =
\$(\op{\rho})$ (see \cite{Chuang-Nielsen,Preskill,SchumacherKrauss} for a 
review of this representation and a derivation of the main formulae). 
The trace preserving nature of the superoperator defines a constraint 
for the operators $\op{E}_k$:
\begin{equation}
\sum_k \adjop{E}_k \op{E}_k = \op{I},\label{sumEE=1}
\end{equation}
where $\op{I}$ is the identity operator. There is no other constraint on 
these operators, although it is worth noting that this constraint boils
down to plain unitarity if there is only one operator. 
There is a nice physical interpretation for the $\set{\op{E}_k}$ operators: 
One can think equation (\ref{eq:opsum}) as corresponding to a process 
where the state $\op{\rho}$ is converted  randomly into the state
$\frac{\op{E}_k\op{\rho}\adjop{E}_k}{\tr\left(\op{E}_k
\op{\rho}\adjop{E}_k\right)}$ with a probability 
$\tr \left(\op{E}_k\op{\rho}\adjop{E}_k \right)$ (in this sense, these 
operators are quantum jump operators). It is important to 
notice that the set of the $\op{E}_k$ corresponding to a given non-unitary 
evolution 
is not unique: For example, if $U_{kl}$ are the elements of a unitary matrix, 
we can define the operators $\op{E}_l'$ as
\begin{equation}
\op{E}'_l = \sum_k U_{kl} \op{E}_k,
\end{equation}
and show that the set $\set{\op{E}_l'}$ generates the same evolution as the set
$\set{\op{E}_l'}$. 

We will use the operator sum representation to define a specific model 
to introduce decoherence in our system. We want our model to correspond, 
in the continuum limit (where $N\rightarrow\infty$) to a diffusive environment
having similar effects than the ones present in the well studied Brownian 
motion model \cite{QBM}. For this, we will assume that the temporal 
evolution is divided in two steps: a unitary step, where the density matrix
evolves unitarily as in (\ref{unitarymap}); and a dissipative
step, where the density matrix evolves by a map whose 
operator sum representation is of the form: 
\begin{equation}
\op{\rho}' = (1-\alpha) \op{\rho} +
\frac{\alpha}{2} (\op{D}(\Delta q,\Delta p) \op{\rho} \adjop{D}(\Delta q,
\Delta p) +
                  \adjop{D}(\Delta q,\Delta p) \op{\rho} \op{D}(\Delta q,
\Delta p)).
\end{equation}
where $\alpha$ is a real number between $0$ and $1$ measuring 
the strength of the
coupling to the environment and $\op{D}(\Delta q,\Delta p)$ is the 
displacement operator defined in (\ref{eq:D_operators}), which 
from now on will be denoted simply as $\op{D}$. 
Notice the constraint (\ref{sumEE=1}) is
automatically satisfied because $\op{D}$ is unitary. In fact, in the 
above equation the three terms
appearing in the operator sum representation are:
\begin{eqnarray}
\op{E}_0 & = & \sqrt{(1 - \alpha)} \op{I}       \\
\op{E}_1 & = & \sqrt{\frac{\alpha}{2}} \op{D}   \\
\op{E}_2 & = & \sqrt{\frac{\alpha}{2}} \adjop{D},
\end{eqnarray}
which are normalized in such a way that (\ref{sumEE=1}) holds. 

The above operator sum representation provides an intuitive interpretation 
for the evolution: the density matrix is first evolved with the unitary
quantum map (the unitary step). Then, three things can happen: i) with 
probability $1-\alpha$ the state does not change; 
ii) with probability $\frac{\alpha}{2}$ the state is displaced in 
one direction in phase space; iii) with probability $\frac{\alpha}{2}$ 
the state is displaced in the opposite direction in phase space. 
As the probabilities of both displacements are equal, a localized state 
does not drift in phase space as a consequence of this evolution. The 
net effect is to smear the state in phase space in the direction of the 
displacement operator. Thus, this nonunitary map is a discrete model for a 
diffusive process. The diffusive superoperator can be made a more 
efficient by using not just a single displacement operator but rather a 
sum of many terms. Thus, we will consider a more general model where
\begin{equation}
\op{\rho}' = (1-\alpha)\op{\rho} + \frac{\alpha}{2M} \sum_{n=1}^{M} (
\op{D_n} \op{\rho} \adjop{D_n} + \adjop{D_n} \op{\rho} \op{D_n} )
\label{diss_map}
\end{equation}
where $\op{D_n}=\op{D}(\Delta q_n,\Delta p_n)$ for some displacements
$\Delta q_n$ and $\Delta p_n$. In particular, it is simplest to 
consider all displacements along the same direction 
(i.e. $\Delta q_n=n\Delta q$
and $\Delta p_n=n\Delta p$). In such case, we can get the following 
formula for the evolution of the matrix elements of $\op{\rho}$
(in the basis of eigenstates of $\op{D_n}$): 
\begin{equation}
\frac{\bra{i'}\op{\rho}'\ket{i}}{\bra{i'}\op{\rho}\ket{i}} = 
1 - \alpha \big[ 1-   \cos(\frac{\pi(i-i')}{N}(M+1))
    \frac{\sin(\frac{\pi(i-i')}{N}M)}{M \sin(\frac{\pi(i-i')}{N})}\big]
\label{diss_action}    
\end{equation}
From this equation we see that the effect on the density matrix 
in this representation is clear: 
the diagonal elements are unaffected by the non--unitary evolution 
while the non-diagonal elements are suppressed by a factor that
rapidly decays with the distance to the diagonal. In the continuous case this
suppression is gaussian \cite{Haake} but here, due to the discreteness of the 
Hilbert space the suppression has the shape of a diffraction-like kernel.
Using the above expression it can be shown that if $M = N$, the net 
effect of one step is to completely wipe out all the non-diagonal elements, 
leaving a diagonal density matrix.

The diffusive character of this non--unitary map is even clearer if we 
represent the quantum state in phase space using the Wigner functions. As 
we are dealing with a finite dimensional Hilbert space we should 
use a discrete version of the ordinary Wigner function which is well 
adapted to the finite phase space structure \cite{DiscreteWigner}. 
The definition and main properties are given in the Appendix and
more in detail in Ref.\cite{Nos:Wigner}. In this representation the Wigner
function of the density operator is a real array defined on a $2N\times 2N$
lattice in phase space which shares many of the well known properties of the
continuous Wigner function. The action of the diffusive map given in equation
(\ref{diss_map}) in the Wigner representations is
\begin{equation}
W'(q,p)  = (1 - \alpha) W(q,p) + \frac{\alpha}{2M} \sum_{n=1}^{M} 
 [ W(q+2n\Delta q, p+2n\Delta p) + W(q-2n\Delta q, p-2n\Delta p)] 
\label{map_wig}
\end{equation}
This expression shows clearly that the diffusive step smears the Wigner
function in directions specified by $\Delta p, \Delta q$ (the factor 
of $2$ present
in the above equation is due to the fact that the phase space is an array of
size $2N\times 2N$ rather than $N\times N$). More general diffusion
models involving mixtures in different directions in phase space, or even
over whole areas where diffusion occurs are similarly represented.

\section{Results}

In this section we present and discuss our results. They concern two 
separate but connected issues: First, we will examin the behavior of
entropy as a function of time focusing both on understanding the role
of the different mechanisms that make entropy to grow and determining 
the dependence of the rate at which it grows on the parameters of our
model (like the strength of the coupling, etc). Second, we analize the 
issue of the correspondence between quantum and classical dynamics. 
The basic tool we used for our studies is a highly efficient code to 
evolve the density matrix which makes good use of fast Fourier transform
routines. With this, running on a PC, one can easily compute the 
density matrix for Hilbert spaces with dimension of about $N=2000$ (larger
machines would be required to go above that limit but there is no good
reason to do that, see below). The source code is available 
from the authors. 

\subsection{Rate of entropy production}

\begin{figure}
\vspace{0.7cm}
\center
\epsfxsize=8.6cm
\epsfbox{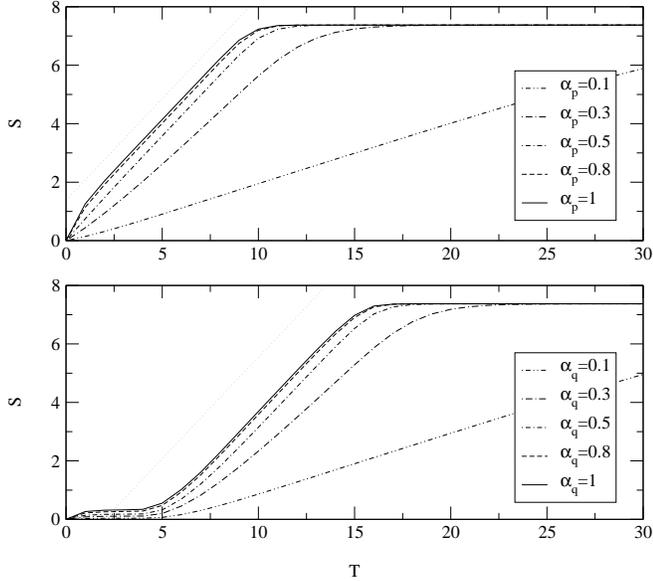}
\vspace{0.25cm}
\caption{Linear entropy growth for the baker's map with N=1594 
and different values for $\alpha$. For the top graph diffusion 
is along momentum and for the bottom graph along position.}
\label{graf:lin_ent1}
\end{figure}

For a variety of initial states we studied the evolution of the density matrix under the dissipative map and computed the linear entropy 
$S=-\log(Tr(\rho^2))$.  
We did that both for the baker's and the Harper's map. 
The diffusion mechanism is as in (\ref{diss_map}) and in the plots $\alpha_p$
labels diffusion along the momentum axis ($\op{D}= \op{\cal V}$ ) while
$\alpha_q$ labels diffusion along the coordinate ($\op{D}= \op{\cal U} $)
Our results clearly establish that in both cases, the rate of entropy 
production becomes {\sl independent} of the parameter $\alpha$ provided
its value is above a certain threshold (that is close to $0.4$). 
This is seen in Figures \ref{graf:lin_ent1} where the plots show the
entropy as a function of time for two types of diffusive environments
(where diffusion is along either position or momentum). 
It stands out from this graph that all of the curves with $\alpha$ greater 
than $0.4$ have a linear regime with the same slope. 
This slope turns out to be equal to $\ln 2$, 
the Lyapunov exponent for the baker's map.

\begin{figure}
\vspace{0.7cm}
\center
\epsfxsize=8.6cm
\epsfbox{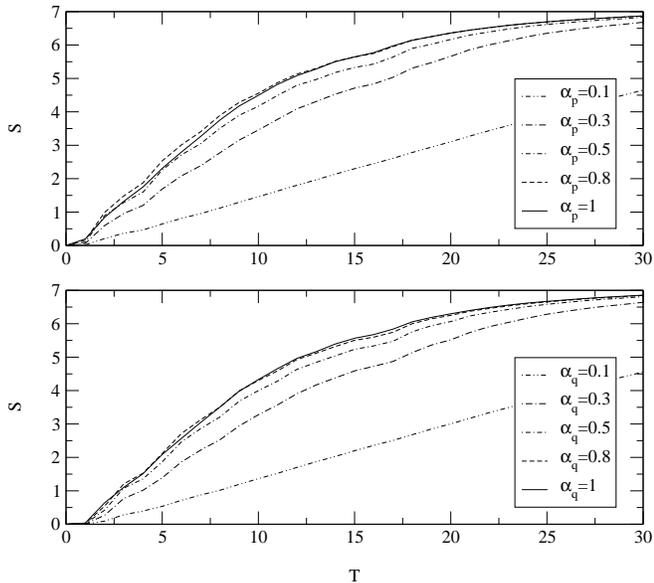}
\vspace{0.25cm}
\caption{Same as Fig \ref{graf:lin_ent1} but for Harper's map ($\gamma$ = 0.45)}
\label{graf:lin_ent2}
\end{figure}

Similar results are shown in Figure \ref{graf:lin_ent2}, for which the
unitary propagation was provided by the Harper map. Even though in this case there is not a well 
defined linear regime (at least not so well defined as in the case of 
the baker's 
map), it is clear that the rate of entropy production becomes independent
of $\alpha$. Thus, in this regime entropy is produced due to the coupling
with the environment but the rate becomes independent of the 
coupling strength. 

This is one of the main results of the paper, that substanciates the conjecture
that was originally put forward in \cite{ZurekPaz94} and later tested 
numerically in various works \cite{MP00,Sarkar,Pattan}.

\begin{figure}
\vspace{0.7cm}
\center
\epsfxsize=8.6cm
\epsfclipon
\epsfbox{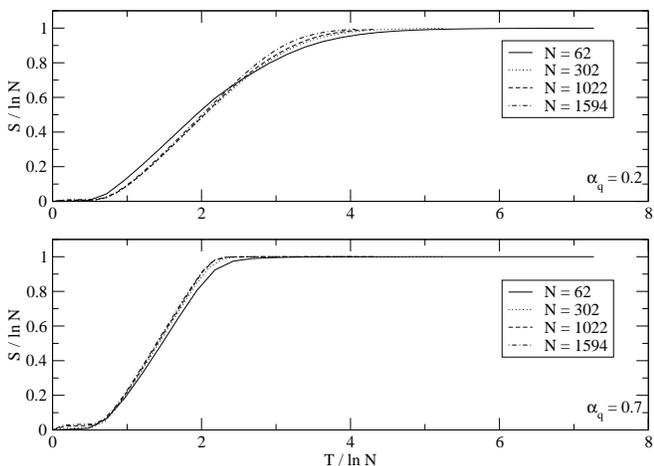}
\vspace{0.25cm}
\caption{Scaling of the linear entropy growth with dimension of the Hilbert 
space of the system with diffusion on the position axis. The strength of the 
diffusion is kept constant by smoothing over a constant fraction of the 
system's phase space. In relationship with Figures \ref{graf:scl1} and 
\ref{graf:scl2} both axis have been rescaled by $\log(N)$}
\label{graf:scl1}
\end{figure}

\begin{figure}
\center
\epsfxsize=8.6cm
\epsfclipon
\epsfbox{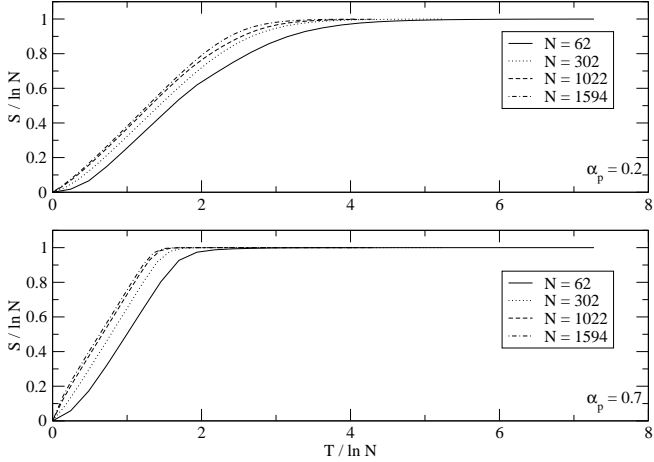}
\vspace{0.25cm}
\caption{Scaling of the linear entropy growth with dimension of the Hilbert 
space of the system with diffusion on the momentum axis. The strength of the 
diffusion is kept constant by smoothing over a constant fraction of 
the system's phase space. In relationship with Figures \ref{graf:scl1} 
and \ref{graf:scl2} both axis have been rescaled by $\log(N)$.}
\label{graf:scl2}
\end{figure}

We have also studied the dependence of the entropy growth as a function 
of $N$ (the dimensionality of the Hilbert space). This was done both for 
small values of $\alpha$ and in the regime where the rate of entropy 
production is independent of the 
coupling. These results are shown in Figures \ref{graf:scl1} and 
\ref{graf:scl2}. In both cases we have found that 
- when appropriately rescaled - the entropy growth is quite independent on the
value of $N$ and tends to a universal curve in the limit $N \to \infty$ 
(as $1/N$ is the effective Planck constant this corresponds to the 
classical limit). This result suggests that the 
mechanism for entropy growth in quantum maps has a dominant classical 
component with corrections that vanish in the classical limit.

Our results show clearly that above some threshold for $\alpha$, 
the rate of entropy production is determined by the dynamics of the 
unitary quantum map, and in great measure by its classical counterpart). 
Why is this the case? We will argue here that it is
possible to attribute the entropy production to two interconnected processes
whose origin can be better explained by using a phase space representation
for the quantum state. The reason is that in this way we can use some
of the intuition we have about the behavior of the classical system. 
Consider an initial state that is represented by a localized and smooth
phase space distribution (for concreteness we base our discussion on 
the Wigner function, see below). The application of the chaotic unitary map
will distort the state in a way that, at least for short times (times smaller
than $\log(N)$), will be consistent with the classical evolution. In a 
hyperbolic region this will mean that the initial wavepacket will be 
stretched along the unstable manifold and 
contracted along the stable one. In the case of the baker's 
map (and only in this case) these two directions are the same all over 
phase space: the vertical direction -that we have called momentum - is 
stable (contracting) and the 
horizontal one - the coordinate - is unstable (expanding). The rate of 
expansion is equal 
to the Lyapunov exponent which in the baker's map (and only in this case) is
constant all over phase space and is equal to $\ln 2$. However, this 
expansion cannot proceed forever. As the phase space is finite, after 
some time the state will start
experiencing the other essential feature of chaotic dynamics: folding. 
As soon 
as this process starts the quantum and classical evolutions will start
deviating significantly from each other due to the quantum interference 
between the different folds. In the  Wigner function (see below) these 
interferences
appear as rapid oscillations with large negative values  happening in 
between the folds and oscillating in directions parallel to them.
These oscillations are the footprint of the quantum interference between 
the different pieces of the delocalized state. 

The two interconnected effects generated by the diffusive map are: (a) 
the destruction of the interference fringes in the Wigner 
function, (b) the smoothing 
and widening of the regions where the Wigner function is already positive. 
It should be clear that the two effects have the same origin: diffusion, but it
is worth distinguishing between them because they are effective for different
types of states and can be analyzed separately under some special 
circumstances. 
In fact, process (a) affects states which have a quantum nature exhibiting 
important oscillations in the Wigner function. On the other hand, 
process (b) is present even for classical systems (in fact, in a classical 
difussive map process (b) is the only source of entropy). The important point 
is that even if we consider an initial state where only process (b) is 
important (i.e., a classical state), the dynamics of the system will create 
large scale quantum interference which will trigger process (a). 

In the baker's map the two processes can be analyzed separately. Thus, 
if we consider a non--unitary map producing diffusion only along the 
momentum (i.e., $\Delta x=0$ in (\ref{diss_map})) only process
(b) will be relevant. In fact, this is the case because an initial 
phase space distribution will tend to be contracted in the momentum
direction while being expanded in the position direction. Folding will 
start to occur when the state reaches the boundary of the phase space. 
After that, quantum interference will develop between the different 
pieces of the packet and will be characterized by fringes which will
be aligned along the vertical direction. Therefore, the fringes will
not be significantly affected by diffusion. On the contrary, the
effect of diffusion will be to fight against the contraction of the
wavepacket along the momentum direction. One can see that diffusion 
will tend to compensate contraction and the Wigner function will tend
to acquire a critical width along momentum. When this happens, the 
entropy will start growing at a rate fixed by the expansion rate which
is the Lyapunov exponent. This is the scenario originally discussed
in \cite{ZurekPaz94}: the entropy grows at the rate fixed by the
Lyapunov exponent when the Wigner function reaches a critical width
along the stable direction (contraction stops but expansion continues
almost unafected by diffusion, therefore area grows exponentially
which implies that entropy grows linearly). 

The other extreme case is to consider a non--unitary map where difussion
is along the position direction (i.e., $\Delta p=0$ in (\ref{diss_map}). 
In that case, diffusion will not prevent the system from contracting
along the stable direction and therefore will not give rise to a critical 
width in that direction. However, entropy will still grow but it's  
origin is entirely due to process (a). In fact. as soon as the phase space
distribution starts to fold, interference fringes develop. These fringes are 
aligned along the momentum axis (since they correspond, roughly, to 
interference between two horizontal strips separated by a distance $1/2$). 
As the strips are macroscopically separated they will be very sensitive to diffusion along position. Thus, 
if the coupling strenght $\alpha$ is such that the fringes are washed 
out in just one iteration the mechanism will produce a mixture of two
horizontal strips every iteration of the map producing one 
bit of entropy. The rate associated to this mechanism is the rate of 
folding that in the case of the baker's map (and only in this case) is
equal to the global Lyapunov exponent. Thus, the fact that both 
processes (a) and (b) give rise to the same rate is an accident of 
the baker's map. In general the rate will not be the Lyapounov exponent
but a combination of such exponent and the global folding rate. In any
case, the general conclusion which is still valid is that the 
entropy production rate for classically chaotic systems will tend
to be dominated by dynamical properties of the system becoming
independent of the strength of the coupling to the environment. 

It is interesting to present a simple analytic model to support 
the above discussion about entropy production. The simplest case is
the one corresponding to process (b), where difussion is along the 
position direction. To simplify our model we will assume that 
the action of the baker's map is just to transform a momentum 
eigenstate into a superposition of two such states:
\begin{equation}
 \op{B}\ket{k_0} = \frac{1}{\sqrt{2}}(\ket{\frac{k_0}{2}} + 
\ket{\frac{k_0+1}{2}})
\end{equation}
 
(this is a rough approximation for the baker's map that neglects 
diffraction  effects and relative phases but that captures the 
essential folding mechanism
of the baker's action in the large $N$ limit and away from saturation ). 
In this approximation, the unitary evolution turns each momentum 
state into a coherent superposition of two momentum states separated 
by a macroscopic distance $N/2$. According to (\ref{diss_action}), 
the momentum diffusion leaves untouched the diagonal - {\it incoherent} 
- elements for the density matrix of
this state while suppressing, in the
large M limit, the off-diagonal parts by a factor

\begin{equation}
\frac{\bra{i}\op{\rho}'\ket{i+N/2}}{\bra{i}\op{\rho}\ket{i+N/2}} \approx
1 - \alpha 
\end{equation}

With these assumptions it is simple to obtain an equation for the 
entropy as a function of time. Thus, after $t$ iterations the density matrix 
has $2^{t+n-1}$ 
matrix elements with value 
$2^{-t}(1-\alpha)^n$, with $n$ ranging from $1$ to $t$. Using
this, we showed that the linear entropy is
\begin{equation}
S(t) = t\ln 2 - \ln \big[(1-\alpha)^2(1+(2(1-\alpha)^2)^t)-1\big].
\end{equation}

\begin{figure}
\vspace{0.7cm}
\center
\epsfxsize=8.6cm
\epsfbox{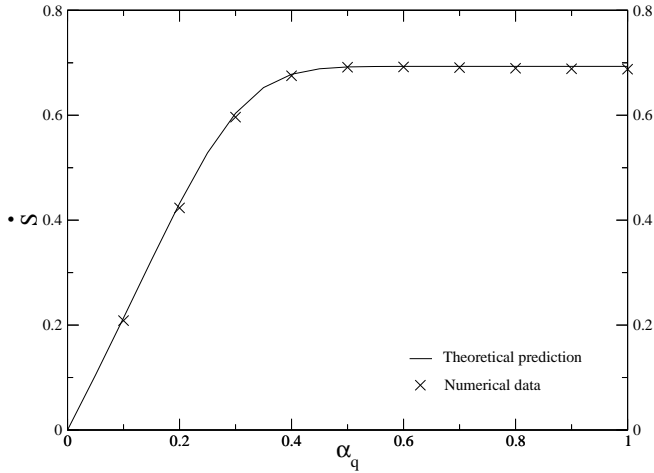}
\vspace{0.25cm}
\caption{$\dot{S}_2$ vs. $\alpha$: Analytic prediction against numerical
results. The analytic line and the numerical data points were not taken from
the same iteration, in order to compensate for the fact that the initial state
in the simulations was not a momentum eigenstate but a coherent state instead,
which takes some iteratios to transform into something approximate to a
momentum eigenstate.}
\label{graf:an_num}
\end{figure}

\begin{figure}
\vspace{0.7cm}
\center
\epsfxsize=8.6cm
\epsfbox{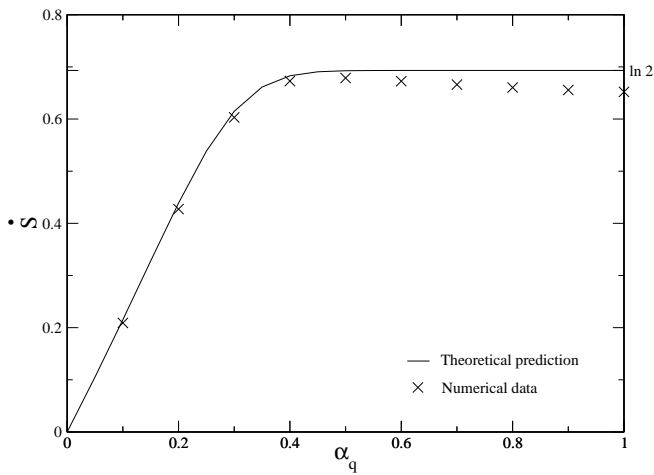}
\vspace{0.25cm}
\caption{$\dot{S}_2$ vs. $\alpha$: Anaylitic prediction against numerical
results some iterations after the ones depicted in graph \ref{graf:an_num}}.
\label{graf:an_num2}
\end{figure}

The second term in this equation corrects the slope in the small $\alpha$
regime. The transition is predicted at a critical value of
 $\alpha_c =1-2^{-1/2}
\approx 0.3 $. For $\alpha > \alpha_c $ the second term becomes intependent of
$t$ and the slope saturates at $ \dot S =\ln 2$. In the weak coupling regime ,
for $\alpha < \alpha_c $ the growth is still linear, but the slope depends on 
the coupling strenght and is given - for long times - by
\begin{equation}
\dot S = -2\ln (1-\alpha) 
\end{equation} 
In this regime the slope of the entropy is then linear with $\alpha$ - and 
independent of the Lyapounov exponent of the map. 
All the predictions of this simple model are confirmed by the numerical
calculations.
In Figure (\ref{graf:an_num}) we display the behavior of
the rate $\dot{S}$ as a function of the coupling strength  $\alpha$ during
the regime of linear growth. The agreement between the simple
model and the exact numerical results is very good, describing  
both coupling regimes very accurately. To see the limitations of the model  
we plot in Figure (\ref{graf:an_num2} the same quantities, but taken at 
later times, closer to the point where the entropy saturates due to the 
finiteness of the model. Here the agreement deteriorates as the assumptions 
of the model break down.

\subsection{Quantum classical correspondence. Wigner function}

We also studied the evolution of the Wigner function for a variety of 
initial states both for the baker's map and the Harper's map. 
In Figures \ref{graf:wig_bak} and \ref{graf:wig_harp} we compare the 
classical (left), unitary (center) and quantum difussive (right) cases
for these two maps. The general features that are observed are the 
following: The unitary map follows the classical evolution for up to 
a time which is of the order of $\log(N)$. After this time quantum 
interference develops between the different pieces of the Wigner 
function. These effects are responsible for the loss of correspondence
between quantum and classical evolution that was discussed in 
\cite{MP01,Berman,HSZ,Volovich}. 
When the effect of the coupling to the environment
is taken into account, it is clear that the correspondence between 
the quantum and the classical evolution is restored and holds for a
much longer time. However, it is worth pointing out that the 
fact that our system is finite (i.e., it has $N$ orthogonal states) implies
that saturation is achieved and the system approaches an equilibrium 
state. This equilibration takes place for times which are a few times $\log(N)$
as seen in the curves for the entropy (thus, as entropy grows linearly, it 
will approach the $\log(N)$ value relatively soon). Therefore, in this
type of system the quantum classical correspondence is restored but 
has a rather short interesting dynamical regime. The nature of the 
final equilibrium state has footprints of the underlying classical 
dynamical system that are evident in the presence of decoherence. This is
not so transparent by analyzing the final state for the Baker map but
it is evident for the Harper's map: the final state with decoherence is
an equilibrium state where occuping uniformily all the available phase 
space with the remarkable exceptions of the regular islands. 

\begin{figure}
\center
\epsfxsize=\textwidth
\epsfclipon
\epsfbox{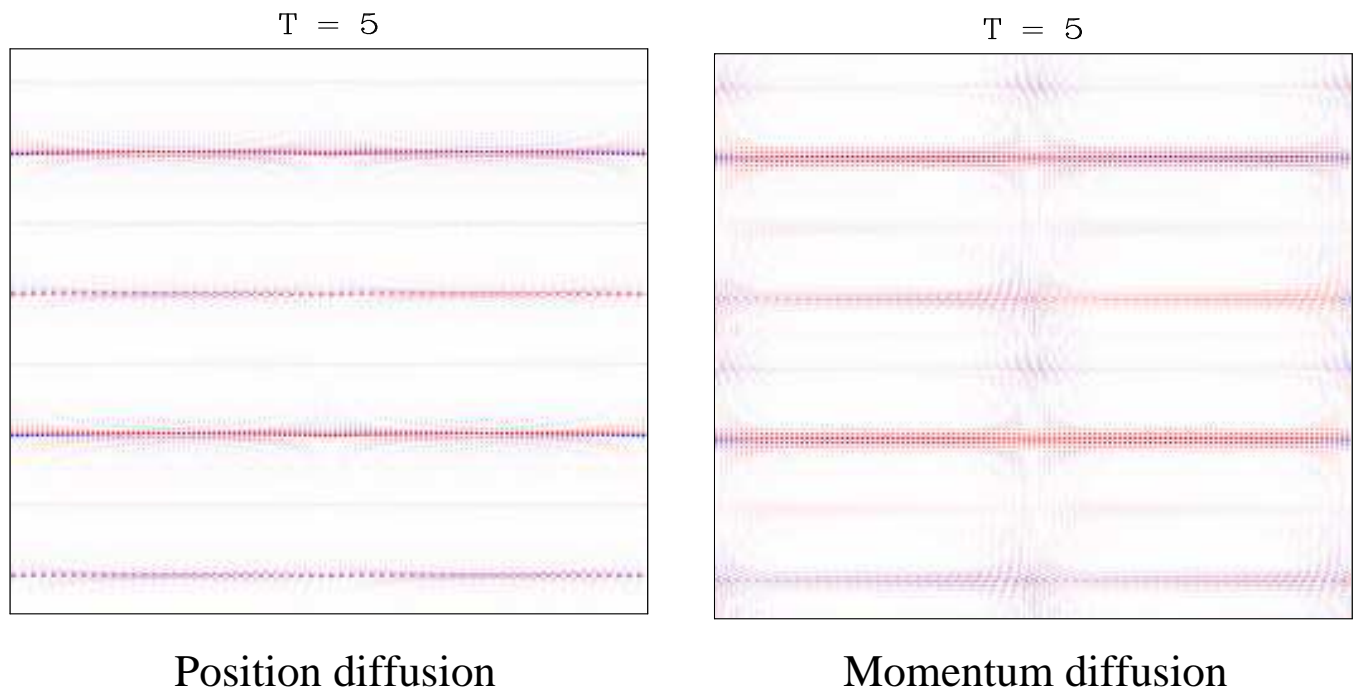}
\vspace{0.25cm}
\caption{Comparison between diffusion in the different manifolds of the baker's
map.}
\label{graf:baker_dqdp}
\end{figure}

By analyzing the Wigner distribution we can study the differences between 
the two processes contributing to the entropy growth that were discussed 
above. Thus, when difussion is along the direction of momentum it tends
to prevent the contraction of the Wigner function but does not affect
the interference fringes. On the other hand, when diffusion is along
the position direction it does not affect the contraction but efficiently
destroys the interference fringes. In Figure \ref{graf:baker_dqdp} 
we show a comparison of the Wigner function for the cases when 
diffusion is in position and momentum. It is clear that in the first 
case the width in momentum is significantly smaller than in the 
second while the interference fringes (that are alligned along momentum)
are washed out much more efficiently. When the diffusion is along the 
stable manifold, the packet width does not shrink in momentum 
and approaches a critical width. Besides, the interference fringes are
more noticeable. In the case of position diffusion there is also 
a critical width that in this case corresponds to the wavelength of the
interference fringes that are no longer efficiently suppressed by 
diffusion (that damps all small wavelength fringes). 

\section{Conclusions}

The main results in this paper come from numerical simulations of open quantum
maps, particularly the baker's and Harper's map coupled to a diffusive
environment. Our simulations provide strong numerical evidence supporting the
conjecture put forward in \cite{ZurekPaz94}: when a quantum system 
whose classical analog is chaotic is subject to decoherence, there is a
regime in which the rate of entropy production is determined by the 
system's dynamical properties (Lyapunov exponents, folding rates, etc) and
becomes independent of the strength of the coupling to the environment. 
Another interesting aspect of our results is the existence of a clear 
scaling for the entropy as a function of the dimensionality (or of the 
effective Planck constant). The scaled curves for the entropy become
independent of $h = \frac{1}{N}$) with quantum corrections that decay as N
grows.  We also found cualitative evidence, looking at the phase space
distributions, that the introduction of diffusion extends the time of
coincidence between the classical and quantum evolutions.
Our results also clarify the origin of entropy production allowing us
to develop new intuitive explanations to understand the origin of the
entropy growth (see \cite{Schack2} for another point of view based
on the use of consistent histories). 
The two processes that are responsible for such growth
(washing out of interference fringes and smoothing of the positive
pieces of the Wigner function) can be studied separately only in the 
case of the baker's map where stable and unstable manifolds have a global
nature in phase space. The differences between the behavior of the entropy
as a function of time for both non--unitary models (diffusion along 
position or along momentum) are illuminating: in both cases there is a 
regime in which the entropy grows linearly with a slope that is equal to 
the Lyapunov exponent (which is also the folding rate). The initial behavior
is different in the two cases: when diffusion is along the position 
direction the entropy stays small initially until the wavepacket reaches
the boundary of phase space and folding becomes effective. Then every 
iteration of the map creates new interference fringes that are rapidly 
destroyed by decoherence. Thus, the dominant process for entropy production
is (a). When diffusion is along the momentum direction, entropy growth starts
at early times. In this case the growth of entropy is produced by process
(b) and is associated with the fact that diffusion tends to balance the
effect of the contraction of the wavepacket along the stable direction 
giving rise to a critical width. In any case, our study 
shows that entropy growth at a rate fixed by the dynamics of the
system rather than the coupling strength to the environment is, 
as first conjectured in \cite{ZurekPaz94} a characteristic
of quantum chaos. 
The differences between this behavior and the one
that characterizes integrable systems (or quantum states localized in 
stability islands like the ones present in Harper's map) are 
quite dramatic (see \cite{Pastawski} for related results). 

This work was partially supported by grants from Ubacyt, Anpcyt, Conicet 
and Fundaci\'on Antorchas. JPP gratefully acknowledges the hospitality
of ITP Santa Barbara where the final version of this manuscript was
completed.

\appendix

\section{Discrete Wigner function}

We use a discrete version
of the Wigner function, which we describe in more detail in \cite{Nos:Wigner}. 
Here we give the basic definitions to make the paper self--contained. 
The Wigner function is usually defined, for a continuous system, as
\begin{equation}
W(q,p) = \frac{1}{2\pi\hbar}\int_{-\infty}^{+\infty} 
                             \bra{q-\frac{y}{2}}\op{\rho}\ket{q+\frac{y}{2}}
                             \cep{\frac{p}{\hbar}y} 
                            \diff{y}.
\end{equation}
This function is the expectation value of the so-called phase space 
point operators:
\begin{equation}
W(q,p) = \tr(\op{\rho}\op{A}(q,p)).
\label{WigDef2}
\end{equation}
The operators $\op{A}(q,p)$ have a number of interesting properties from 
which the properties of the Wigner function can be derived (in fact, they
are hermitian operators, they form a complete basis, etc; see 
\cite{Nos:Wigner}). It turns out that in order to generalize all these 
properties to the discrete case it is necessary to use a phase
space lattice of $2N\times 2N$ points and to define the phase space
operators (in terms of the shift operators used in the body of the 
paper and defined in (\ref{eq:schwinger_ops})) as
\begin{equation}
\op{A}(q,p)=\frac{1}{2N} \op{\cal U}^q \op{R} 
\op{\cal V}^{-p} \exp{i\pi qp/N}.
\end{equation}
Using these discrete phase space point operators the Wigner function
has all the desired properties (it is real, it can be used to compute
expectation values of operators and gives rise to all the correct 
marginal distributions when summed over any set of lines in the phase
space grid). To explicitly compute this function one can use a variety
of formulae among which we found convenient the following expression:
\begin{eqnarray}
W(q_j, p_i) & = & \frac{2}{N^{3/2}} 
           \cem{\twopiN (\frac{q_j p_i}{2} - 2 \flop\floq)} \times \nonumber \\
            &   &  \sum_n \sum_k 
                            \cep{\twopiN (q_j k + p_i n)}
                            \cem{\twopiN (n-\flop)(k-\floq)}
                            \bra{k}\op{\rho}\ket{n}
\end{eqnarray}
On the other hand, it takes a atraightforward calculation to 
show that the $\op{A}(q,p)$ operators transform in the following 
way when subjected to translations (both in their continuous and 
discrete
versions):
\begin{equation}
\op{D}_{(\Delta q, \Delta p)} \op{A}(q,p) \adjop{D}_{(\Delta q, \Delta p)}
                           = \op{A}(q+2\Delta q, p+2\Delta p)
\end{equation}
We can use this property and the definition of the Wigner function 
given in equation \ref{WigDef2} to derive the equation \ref{map_wig}
we use in the paper.

%
%

\onecolumn
\begin{figure}
\center
\epsfysize=20.0cm
\epsfclipon
\epsfbox{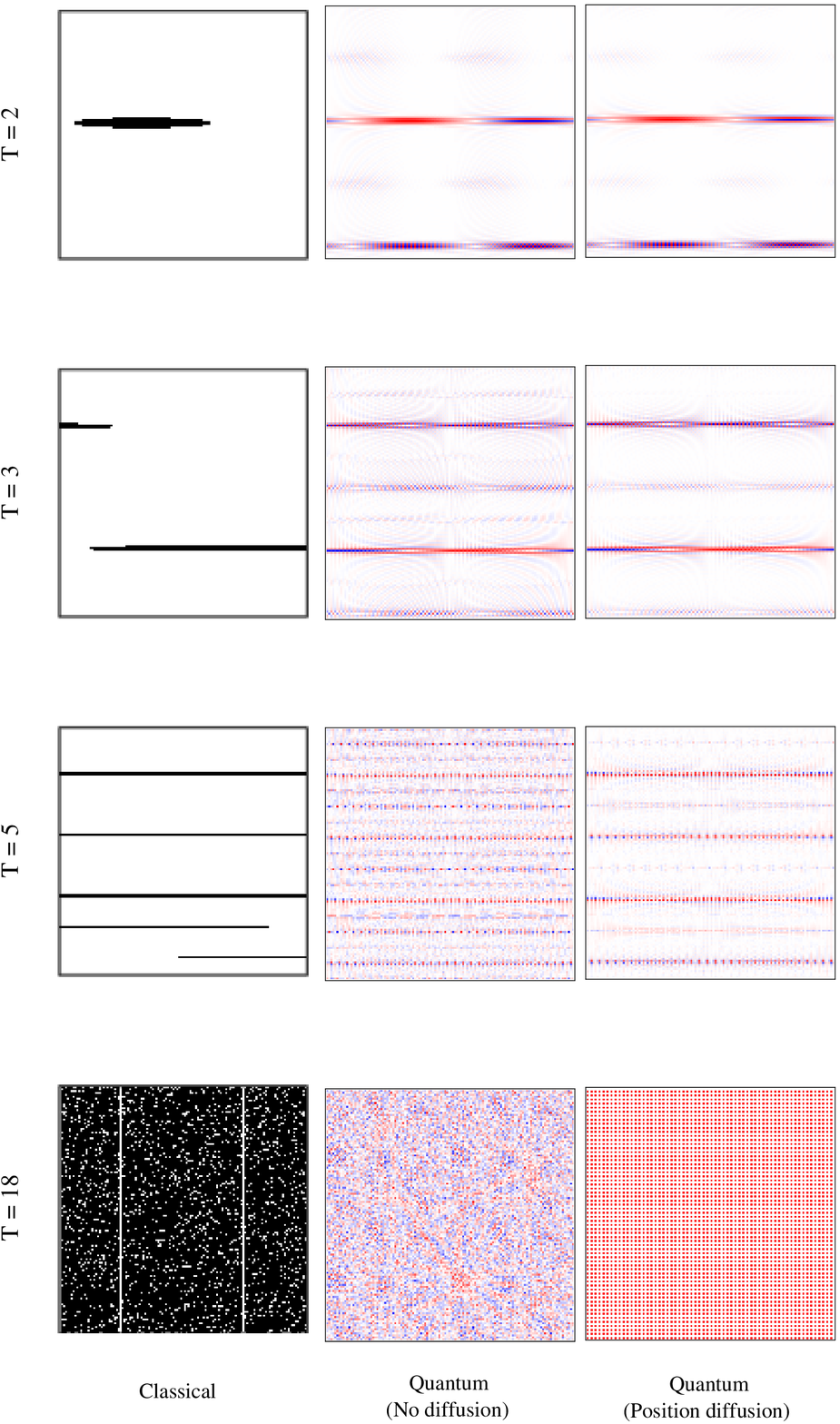}
\vspace{0.25cm}
\caption{Phase space distribution at several steps in the evolution of the
Baker map.  The quantum versions have been computed with $N=62$.}
\label{graf:wig_bak}
\end{figure}

\begin{figure}
\center
\epsfysize=20.0cm
\epsfclipon
\epsfbox{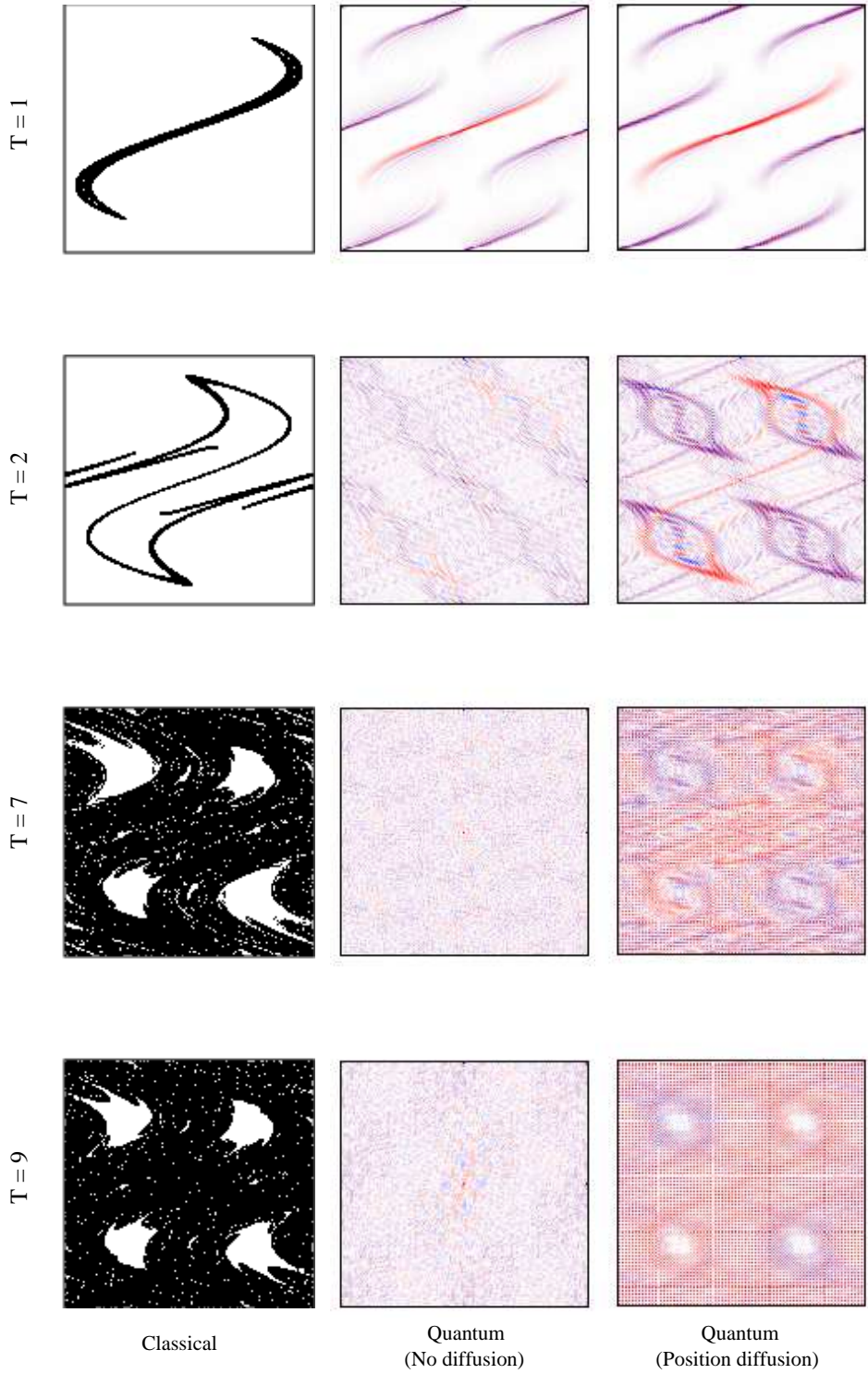}
\vspace{0.25cm}
\caption{Phase space distribution at severals step in the evolution of the
Harper's map. The quantum version has diffusion parallel to the position axis
and $N=62$, $\alpha_q=0.5$, $M=2$.}
\label{graf:wig_harp}
\end{figure}

%
%

\end{document}